\documentclass[prl,twocolumn,showpacs]{revtex4}

\def\>{\rangle}
\def\<{\langle}

\usepackage{graphicx}
\usepackage{amsmath}
\usepackage{amsfonts,amssymb,graphicx,latexsym,color,revsymb}
\begin{document}

\title{High-concurrence steady-state entanglement of two hole spins in a quantum dot molecular}
\author{Song Yang}
\author{Ming Gong}
\author{ChuanFeng Li}
\author{XuBo Zou\footnote{xbz@ustc.edu.cn}}
\author{GuangCan Guo}
\affiliation{Key Laboratory of Quantum Information, University of
Science and Technology of China, Hefei, 230026, People's Republic of
China}

\date{\today}

\begin{abstract}
Entanglement, a non-trivial phenomena manifested in composite
quantum system, can be served as a new type of physical resource in
the emerging technology of quantum information and quantum
computation. However, a quantum entanglement is fragile to the
environmental-induced decoherence. Here, we present a novel way to
prepare a high-concurrence steady-state entanglement of two hole
spins in a quantum dot molecular via optical pumping of trion
levels. In this scheme, the spontaneous dispassion is used to induce
and stabilize the entanglement with rapid rate. It is firstly shown
that under certain conditions, two-qubit singlet state can be
generated without requiring the state initialization. Then we study
the effect of acoustic phonons and electron tunnelings on the
scheme, and show that the concurrence of entangled state can be over
0.95 at temperature $T = 1 $K.
\end{abstract}

%\pacs{03.65.Wj}
\maketitle

\textit{Introduction}- Semiconductor technology toward quantum
information science has opened up the possibility of constructing
scalable quantum devices. As an attractive host for storing quantum
information bit (qubit), electron or hole spins in self-assembled
quantum dot(QD), are most promising for their scalability,
relatively ease of coherent manipulations\cite{ronald} and strong
robustness against relaxation\cite{Elzerman,Heiss}. In the past few
years, significant theoretical and experimental works have been made
towards controlling and entangling quantum dots. These experiments
include the efficient state initialization\cite{emary} and coherent
population trapping of single spin\cite{berezovsky}, the
spin-readout\cite{Elzerman} and single-spin Faraday/Kerr rotations
for single quantum dot spin\cite{Atature,Elzerman}, as well as the
inter-dot coupling in double quantum dots
molecules\cite{Lucio,bayer}. Theoretically, several schemes for
entangling quantum dots have also been proposed \cite{saikin,kolli}.

However, all these schemes are exclusively tailed for electron spins
in self-assembled quantum dot. Due to the longer coherence time
compared with electron spin, hole spin in quantum dot has been paid
more and more attention. A hole-spin state, which is constructed
from a p-type atomic wave function, has many favorable aspects such
as highly-suppressed hyperfine interaction\cite{Brian} and much
smaller tunneling rate\cite{emary1}, in comparison with electron
spin. Recently both experiments have been reported for initializing
single hole spin with high fidelity of 0.99\cite{Brian}, and
creating the coherent population trapping state\cite{Daniel}.

In this letter, we present a scheme to generate high-fidelity steady
state entanglement of two hole in a coupled QDM. Our scheme is based
on spontaneous emission where the coupling between the QDM is
dominated on F\"oster resonant. We show that our scheme is
initial-state independent, and robust against decoherence and
tunneling effect. At $T=1$ K, concurrence higher than 95\% is
possible in state-of-art technology.

\begin{figure}
\includegraphics[width=8.5cm]{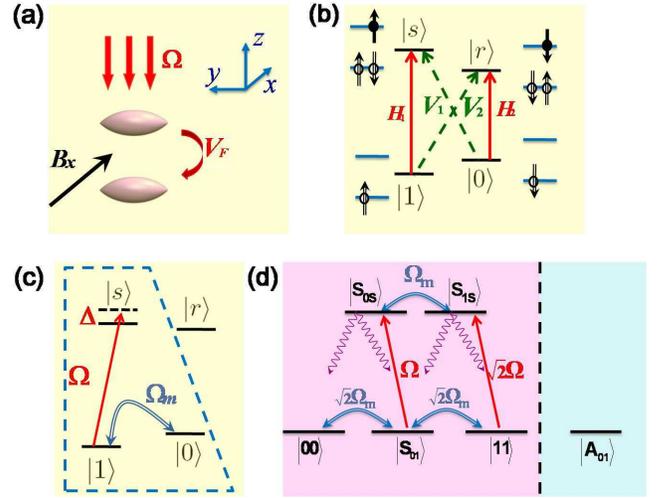}
\caption{(Color Online).(a)Sketch of the two vertically stacked
quantum dots in QDM with Voigt geometry magnetic field.  F\"{o}rster
interaction $V_{F}$ exists between such two resonant QDs. (b) and
(c) Four level scheme illustrating the ground and excited states of
a single self-assembled QD in the Voigt Configuration. (d)The
preparation process of entangled state.} \label{fig2}
\end{figure}

\textit{The model} - The QDM system composes two identity vertical
aligned QDs, where an external magnetic field in the Voigt geometry
\cite{emary} is used to break the degenerate of the energy
level\cite{Kramers}(Fig.\ref{fig2} (a)). Due to the build-in strain
that from lattice mismatch between QDs and the host materials, the
degeneracy of the heavy hole and light hole at $\Gamma$ point is
broken, hence only the heavy hole is only taken into account.
Initially, each QD is dopped with one hole, where the initial spin
polarization is not important in our scheme. The energy level of QD
is shown in Fig. \ref{fig2} (b), the ground hole states are
$|1\rangle = |\Uparrow\rangle$ and $|0\rangle = |\Downarrow\rangle$,
and the excited trion states are $|s\rangle =
|\Uparrow\Downarrow\uparrow\rangle$ and $|r\rangle =
|\Uparrow\Downarrow\downarrow\rangle$, where $\Uparrow(\Downarrow)$
and $\uparrow(\downarrow)$ denote a heavy hole and an electron with
spins along (against) $x$ direction. As the magnetic field is along
$x$ direction, there is an additional Zeeman splitting
$E_{B}^{h(e)}= g_{h(e)}^{*}\mu_{B}B_{x}$ between ground (excited)
states. Because of this splitting, four transitions between ground
states and trion states can be independently addressed by
polarization and frequency selection. Here we can choose $H =
\sigma_{+}-\sigma_{-}$ and $V = \sigma_{+}+\sigma_{-}$.

\textit{Optical pumping protocol} - In generally, the tunneling rate
decreases exponentially when increasing the distance between two
dots, while the Coulomb interaction, including both static dipole
coupling $V_{xx}$ and F\"{o}rster interaction
$V_{F}$\cite{kim,Crooker}, decreases generally $\propto d^{-3}$. So,
it is possible to choose a property distance, where the Coulomb
interaction dominant the interactions between the QDM. The Static
dipole coupling $V_{xx}$ is the bi-trion energy shift, while the
F\"{o}rster interaction between the QDM is a kind of inter-dot
interaction.

Due to the Voigt geometry magnetic field, the energy detuning
between two $H$ type transitions are $\Delta_{H} =
E_{B}^{e}+E_{B}^{h}$, while the energy detuning between $V$ type
transitions are $\Delta_{V} = E_{B}^{e}-E_{B}^{h}$. In our scheme,
we drive a $H$ polarized laser, and the detuning $\Delta_{H}$
enables the transitions $H_{1}$ or $H_{2}$ to be independently
addressed, which has been achieved in experiments with high
fidility\cite{emary}. In the following we choose the transition
$H_{1}$, thus the energy levels of QD employed in our scheme can be
described by three states: $|0\rangle$, $|1\rangle$ and $|s\rangle$
(Fig.\ref{fig2} (c)). The frequency and Rabi frequency with $H$
polarized laser are supposed to be $\omega_{l}$ and $\Omega$
respectively. Additionally, since the direct excitation of the
transition $|0\rangle\leftrightarrow|1\rangle$ is forbidden,
$\Omega_{m}$ can be realized by employing a Raman transition with
large detuning to an auxiliary excited state. The Hamiltonian of the
QDM reads

\begin{eqnarray}
H&=&\sum\limits_{i=1,2}[(\Omega|1\rangle_{i}\langle s|e^{i\omega_{l}
t} +\Omega_{m}|0\rangle_{i}\langle 1|+H.c.)+\omega
|s\rangle_{i}\langle s|]\nonumber\\&+&V_{F}(|1s\rangle\langle
s1|+|0s\rangle\langle s0|+H.c.)+V_{xx} |ss\rangle\langle
ss|.\label{eq:Ham}
\end{eqnarray}

Since Hamiltonian Eq.[\ref{eq:Ham}] is of the symmetry formation, it
is convenient to introduce symmetric state $|S_{ij}\rangle =
\frac{1}{\sqrt{2}}(|ij\rangle+|ji\rangle)$ and anti-symmetric state
$|A_{ij}\rangle = \frac{1}{\sqrt{2}}(|ji\rangle-|ij\rangle)$($i,j =
0,1,s$). With the aid of F\"{o}rster interaction, an energy shift is
generated between symmetric excited states ${|S_{0s}\rangle,
|S_{1s}\rangle}$ and anti-symmetric excited states ${|A_{0s}\rangle,
|A_{1s}\rangle}$.

When the $H$ polarized laser is driven to pump $H_{1}$ transition
with detuning $\Delta = -V_{F}$, the transitions
$|S_{01}\rangle\leftrightarrow|S_{0s}\rangle$ and
$|11\rangle\leftrightarrow|S_{1s}\rangle$ are resonant in the
rotating frame. In the case $\Omega, \Omega_{m}\ll |V_{F}|\ll
V_{xx}$, the populations on bi-trion and anti-symmetric single-trion
states are nearly equal to zero, and can be eliminated
adiabatically. Using the symmetric and anti-symmetric notation we
introduced above, the scheme is reduced to a 6-state system
(Fig.\ref{fig2}(d)). The effective Hamiltonian can be written as
\begin{eqnarray}
H_{\mathrm{eff}}&=& \sqrt{2}\Omega |11\rangle\langle S_{1s}|+\Omega
|S_{01}\rangle\langle S_{0s}|+
\Omega_{m} |S_{0s}\rangle\langle S_{1s}| \nonumber \\
&+& \sqrt{2} \Omega_{m} |00\rangle\langle S_{01}| + \sqrt{2}
\Omega_{m} |S_{01}\rangle\langle 11| + h.c.. \label{eq:Hamsy}
\end{eqnarray}

Then we take the lifetime of trion states into account. The photon
emission occurs via an decay of the state $|s\rangle$ into
$|1\rangle$ with $\Gamma_{1}$ or into $|0\rangle$ with $\Gamma_{0}$.
The total spontaneous rate is assumed to be
$\Gamma=\Gamma_{0}+\Gamma_{1}$. Then we derive a master equation
within a Markovian process,
\begin{eqnarray}
\dot{\rho}=-i[H_{\mathrm{eff}},\rho]+\sum \limits_{i}(
L\rho L^{\dagger}-\frac{1}{2} \{L^{\dagger}L\rho\}_{+}),
\label{eq:ME}
\end{eqnarray}
where $L_{1}=\sqrt{\Gamma_{0}}(|00\rangle\langle
S_{0s}|+\frac{1}{\sqrt{2}}|S_{01}\rangle\langle S_{1s}|)$,
$L_{2}=-\sqrt{\Gamma_{0}/2}|A_{01}\rangle\langle S_{1s}|$,
$L_{3}=\sqrt{\Gamma_{1}}(|11\rangle\langle
S_{1s}|+\frac{1}{\sqrt{2}}|S_{01}\rangle\langle S_{0s}|)$,
$L_{4}=\sqrt{\Gamma_{1}/2}|A_{01}\rangle\langle S_{0s}|$.

\textit{Steady state entanglement} -The basic preparation cycle in
our scheme works as follows. The lasers with Rabi frequency
$\sqrt{2}\Omega_{M}$ produce the transitions
$|00\rangle\leftrightarrow|S_{01}\rangle\leftrightarrow|11\rangle$.
Attribute to the F\"{o}rster interaction, energy splitting generates
between the symmetric and anti-symmetric states. By tuning the laser
frequency to resonant with the symmetric single-trion states, the
transitions $|S_{01}\rangle\leftrightarrow|S_{0s}\rangle$ and
$|11\rangle\leftrightarrow|S_{1s}\rangle$ are created. Thus the
lasers couple the three ground states $|00\rangle$, $|S_{01}\rangle$
and $|11\rangle$ to the single-trion states $|S_{0s}\rangle$ and
$|S_{1s}\rangle$, and leave the entangled state $|A_{01}\rangle =
\frac{1}{\sqrt{2}}(|10\rangle-|01\rangle)$ decoupled from them. On
the other hand, spontaneous radiation performs from $|s\rangle$ to
$|0\rangle$ and $|1\rangle$, which is dissipation from single-trion
states to subspace $M_{1}=\{|00\rangle,|S_{01}\rangle,|11\rangle\}$
and $M_{2}=\{|A_{01}\rangle\}$. If the excited states decay into
$M_{2}$, the process is terminated when one trion is dissipated; if
they decay into $M_{1}$, it will go through the cycle again.
Therefore, after a period of time, the trions are dissipated and the
system will go into a steady state $|A_{01}\rangle$, which is the
maximum entangled state we require.

\begin{figure}
\includegraphics[width=8.5cm]{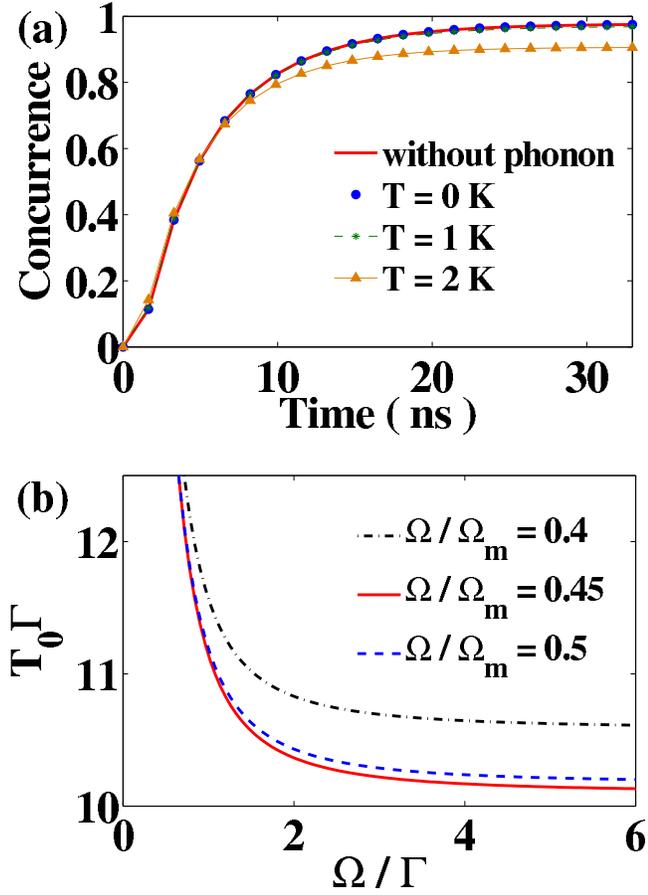}
\caption{(Color Online). (a) Concurrence of the entangled state as trion-phonon
effect is taken into account. The relative parameters: $\Omega = 20$ $\mu$eV,
$\Omega_{m} = 9$ $\mu$eV,  $\Gamma = 1.2$ $\mu$eV.
(b) The characteristic time $T_{0}$ as a function of Rabi frequencies $\Omega$ and
$\Omega_{m}$, as well as total spontaneous rate $\Gamma$.}
\label{fig3}
\end{figure}

For a potential experimental system, we can choose a QDM including
two resonant QDs. The wave functions for electrons and holes in each
QD is supposed to have a Gaussian form $\phi_{e/h}\sim
\exp[-{(x^2+y^2)\over l_{\parallel}^2} - {z^2\over 2 l_{\perp}^2}]$,
where the characteristic lengths $l_{\parallel(\perp)} =
\sqrt{\hbar/(\omega_{\parallel(\perp)} m_{e/h})}$. We assume that
$l_{\parallel e}=4.4$ nm, $l_{\parallel h}=4$ nm, $l_{\perp}=1$ nm.
The F\"{o}rster interaction \cite{Emil} read as $|V_{F}| =
\frac{e^{2}|a|^{2}}{4\pi\varepsilon
d^{3}}(\frac{l^{2}}{l_{e}l_{h}})^{2}F(\frac{d}{l})$, where
$\varepsilon$ is the dielectric constant, $d$ is the distance
between the QDM, $l^{2}= 2/(1/l_{e}^{2}+ 1/l_{h}^{2})$ and $a\simeq
\hbar/\sqrt{2 m_{e}E_{g}}$\cite{PYyu}. In the case $l_{e}, l_{h}\gg
l_{z}$,
$F(x)=\frac{x^{3}}{2\pi}\int_{0}^{1}dt\frac{1-2\nu}{\sqrt{1-t^{2}}}
\exp[-\nu]$, where $\nu = \frac{x^{2}t^{2}}{2(1-t^{2})}$. In our
scheme, we set the distance between two dots in QDM is $9.5$nm and
$|a|=1.6$ nm, We can estimate $V_{F} = - 0.2$ meV and $V_{xx} = 3$
meV. If the $x$-direction magnetic field is $B_{x}= 1$ T, and
g-factors are $g_{h}^{*} = -0.29$ and $g_{e}^{*} = -0.46$, the
Zeeman splitting of trion and ground states are about $E_{B}^{e} =
-27.78$ $\mu$eV and $E_{B}^{h} = -17.94$ $\mu$eV. Thus the detuning
between $H_{1}$ and $H_{2}$ is $|\Delta_{B}| = 45.72$ $\mu$eV. Due
to the requirement of frequency selection, the Rabi frequency of
laser $\Omega$ should satisfy $\Omega\ll|\Delta_{B}|$, and $\Omega =
20$ $\mu$eV is suitable.

Fig.\ref{fig3}(a) shows the entanglement of the two hole spins in
steady state measured by concurrence $\mathcal{C}$ as a function of
time\cite{Wootters}. The Red solid line represents the case only
with spontaneous radiation, which shows a near unity concurrence at
$T = 0$ K after a period of time. Here we suppose that
$\Gamma_{1}=\Gamma_{2}=\Gamma/2$. The input density matrix we
consider is given by $\rho_{i} =
\frac{1}{4}(|00\rangle\langle00|+|S_{01}\rangle\langle
S_{01}|+|A_{01}\rangle\langle A_{01}|+|11\rangle\langle11|)$. After
a period of time, the output state ideally generates as
$|\Phi_{f}\rangle=|A_{01}\rangle$. If the evolution time
$t\rightarrow\infty$, the density matrix
$\rho\rightarrow|A_{01}\rangle \langle A_{01}|$. It is illustrated
in Fig.\ref{fig3}(b) that the character time $T_{0}$, which
expresses that the time for achieving the steady
state\cite{Michaeli}, depends on spontaneous radiation rate $\Gamma$
and the Rabi frequency of pumping field $\Omega$.  If the rate of
spontaneous emission is fixed, the character time can be shorten as
Rabi frequency of laser field increases, and finally saturates
approximal to a value $T_{0}\sim 10/\Gamma$, which is ten times that
of the lifetime of trion state. The optimal characteristic time is
appeared when we tune the coupling $\Omega_{m}$ to satisfy
$\Omega_{m} = 0.45\Omega$(shown in Fig.\ref{fig3}(b)).

For a real QDM, the rate of the spontaneous radiation from trion
state to electronic state is about 1.2 $\mu$eV, therefore $T_{0}$
has a value about $5.5$ ns.

\textit{The effect of phonon interaction} - In Fig.\ref{fig2}, we
have taken exciton as the auxiliary state, which is vulnerable by
the vibrational modes of the surrounding phonons. The interaction
between acoustic phonons and excitons may be mediated by deformation
potential coupling and piezoelectric coupling. Thus the phonon
coupling matrix element\cite{Krumm} is
\begin{equation}
g_{\mathbf {q},j}=e^{{i\mathbf{q}\cdot
\mathbf{d_{j}}}}[M_{q,j}^{e}\rho_{e}(\mathbf{q}) -M_{q,j}^{h}
\rho_{h}(\mathbf {q})],\label{eq:MEph}
\end{equation}
where  $M_{q,j}^{e(h)}=\sum_{\mathbf{q}}\sqrt{\frac{\hbar}{2\mu |q|
V c_{s}}}(|q|D_{e(h)}+ i P_{\mathbf{q}})$, $\rho_{e/h}(\mathbf
{q})=\int d^{3}r|\phi_{e/h}|^2e^{i\mathbf{q}\cdot\mathbf{r}}$.
Following the Markovian approximation\cite{Erik}, the master
equation of the density matrix in the interaction picture with
respect to $H$, may be reduced into a Lindblad form
\begin{eqnarray}
\dot{\rho}&=&\sum_{i}J(\omega_{i})[(N_i+1)D[P_i]\rho + N_iD[P^{\dag}_{i}]\rho],
\label{eq:master}
\end{eqnarray}
where $D[P]\rho = P\rho P^{\dag}-{\frac{1}{2}}\{P^{\dag}P\rho\}_{+}$
is the decay operator of phonon effect, and $N_i=[\exp(\omega_i/
k_{B}T)-1]^{-1}$. $J(\omega_{i})$ denotes the phonon spectral
density, and there are two kinds of $J(\omega_{i})$ in our model
which can be written as
\begin{equation}
J_{\pm}(\omega) = \int d\Omega (1\pm \mathrm{sinc}(\frac{\omega
d}{c_{s}}))[\mathbb{G}_{d}(\omega)+\mathbb{G}_{p}(\omega)],
\end{equation}
where $\mathbb{G}_{d}(\omega)=\frac{\omega^{3}}{8\pi^{2}\mu
c_{s}^{5}}(D_e \varrho_{e}-D_h \rho_{h})^2$,
$\mathbb{G}_{p}(\omega)=\frac{\omega|P_{\mathbf{q}}|^{2}}{8
\pi^{2}\mu c_{s}^{3}}(\varrho_{e}-\rho_{h})^2$. Here the
piezoelectric coupling is $P_{\mathbf{q}} = \frac{1}{4}\sin \theta
M_{p} \sqrt{9+7 \cos2\theta-2\cos4\varphi \sin^{2}\theta}$, in which
$M_{p}$ denotes piezoelectric constant\cite{taka}. Combining the
phonon effect into the Master equation Eq. (\ref{eq:ME}), we will
get numerical results of the concurrence shown in Fig.\ref{fig3}(b).
As temperature increases, the concurrence of entangled state
decreases. The parameters for the phonons is taken from Ref.
\cite{Krumm}.

\begin{figure}
\includegraphics[width=8cm]{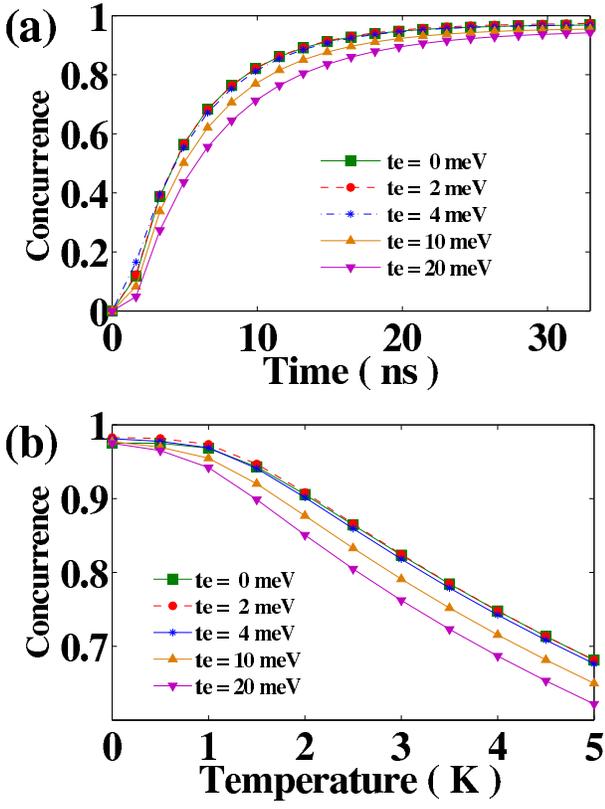}
\caption{(Color Online). Concurrence of the entangled state for different
electron tunneling rates: (a) as a function of time at $T = 1$K.
(b) as a function of temperatures.}
\label{fig4}
\end{figure}

\textit{The effect of tunneling effect} - The tunneling effect is
assumed to be much smaller than the F\"{o}rster
interaction\cite{Nazir} in above discussion. This assumption can be
technically accomplished by increasing the distance between double
dots but simultaneously the F\"{o}rster interaction is suppressed.
It is demonstrated that the tunneling effect of hole is less than
that of electron by one or two orders of magnitude\cite{emary1}
which guarantees the stability of the hole-included ground state.
Thus, we primarily consider the influence of electron tunneling on
our scheme. Using standard WKB method, the tunneling rate can be
estimated to $t_{e}\simeq \frac{2e}{\pi} \sqrt{8 V'_{e} \omega}
\exp[-\frac{16 V'_{e}}{3\omega}]$, with $\omega =
4\sqrt{2V'_{e}}/d$. If the case is $d = 9.5$ nm, $V'_{e} = 680$ meV,
and $m_{e} = 0.067$, the tunneling rate is $t_{e} = 1.9$ meV, which
can not be omitted compared to the F\"{o}rster coupling
$|V_{F}|=0.2$ meV.

The exciton we discuss above in trion $|s\rangle$ is intra-dot
exciton(an electron-hole pair lies in the same dot), such as
$|e_{1}^{\dag}h_{1}^{\dag}\rangle$ or
$|e_{2}^{\dag}h_{2}^{\dag}\rangle$, where
$e_{i}^{\dag}$($h_{i}^{\dag}$) denotes that one electron(hole)
generates in the $i$th dot. Due to the electron tunneling effect,
the trion might include inter-dot exciton
($|e_{1}^{\dag}h_{2}^{\dag}\rangle$ or
$|e_{2}^{\dag}h_{1}^{\dag}\rangle$), and we denote this kind of
trion as $|t\rangle$. Here, we do not consider the spontaneous
radiation caused by inter-dot exciton. Therefore the effect of
inter-exciton gives a new contribution to the initial Hamiltonian
describing by Eq. (\ref{eq:Ham}) as

\begin{eqnarray}
 H_{t}&=&\sum\limits_{i=1,2}(\omega_{t}|t\rangle_{i}\langle t|+t_{e}(|s\rangle_{i}\langle
 t|+H.c.))\label{eq:Hamte}.
\end{eqnarray}

We move the combined hamiltonian including tunneling effect into an
interaction picture with respect to
$\sum\limits_{i=1,2}(\omega+V_{F})(|s\rangle_{i}\langle s| +
|t\rangle_{i}\langle t|)$,  and proceed by transforming the single
exciton part of hamiltonian into a new basis as $|\psi_{1(3)}\rangle
= \cos\theta |S_{0(1)x}\rangle - \sin \theta |S_{0(1)t}\rangle $,
$|\psi_{2(4)}\rangle = \sin\theta |S_{0(1)x}\rangle + \cos\theta
|S_{0(1)t}\rangle$, where $\theta = -
\frac{1}{2}\mathrm{arccot}(\frac{\delta}{2 t_{e}})$ and the detuning
$\delta = V_{F}+\omega - \omega_{t}$. These two eigien states
$|\psi_{1}\rangle$ and $|\psi_{3}\rangle$ are degenerated at energy
$E_{1} = \frac{1}{2}(-\delta+\sqrt{4 t_{e}^{2}+\delta^{2}})$, while
$|\psi_{2}\rangle$ and $|\psi_{4}\rangle$ are degenerated at energy
$E_{2} = \frac{1}{2}(-\delta-\sqrt{4 t_{e}^{2}+\delta^{2}})$. If we
select the frequency of pumping laser as $\omega_{l} = \omega +
V_{F} + E_{1}$, and guarantee  the condition $\Omega \ll |E_{1} -
E_{2}|$, the effective Hamiltonian becomes as

\begin{eqnarray}
H_{\mathrm{eff}} &=&
\Omega_{m}(\sqrt{2}|00\rangle\langle
S_{01}|+\sqrt{2}|S_{01}\rangle\langle11| + |S_{0s}\rangle\langle
S_{1s}|)\nonumber\\ &+&
\Omega\cos\theta
 (\sqrt{2}|11\rangle\langle\psi_{3}|+
|S_{01}\rangle\langle\psi_{1}|) + h.c. .
\label{eq:Hamteeff}
\end{eqnarray}

We consider the influence of both tunneling effect and phonon
interaction and follow similar method as in Eq.(\ref{eq:MEph}). The
concurrence of hole spin entangled state as a function of time at $T
= 1 $K is shown in Fig.\ref{fig4} (a). We can find that the
concurrence is beyond $95\%$, and the phonon-exciton process might
be suppressed by decreasing electron tunneling rate. When the
electron tunneling rate is slow as $t_{e}\ll |\omega-\omega_{t}|$,
the Eq. (\ref{eq:Hamteeff}) can be reduced to Eq. (\ref{eq:Hamsy}).
It indicates that the small electron tunneling does nothing more
than an energy shift of the single exciton states whose effect can
be offset by tuning the Rabi frequency of the external laser field.
Fig.\ref{fig4} (b) illustrates the concurrence of the steady state
as a function of experimental temperature $T$. The high concurrence
of stationary entangled state is much less influenced when
decreasing the temperature, and the concurrence is more susceptive
to temperature if electron tunneling rate increases. In our scheme,
the electron tunneling has a value of $t_{e}\sim 2$ meV, which is
much smaller than the energy gap between inter-exciton and
intra-exciton($\sim 20$ meV). Thus, the hole spin entangled state
remains robust even if electron tunneling effect is considered.

\textit{Conclution} - To sum up, we have shown that a stationary
entangled state on spins with high concurrence can be prepared in a
quantum dot molecular by technically designing the spontaneous
dispassion processes. The hole spin for its small inter-dot
tunneling rate is more suitable to encode qubit compared with
electron spin in our scheme. We also discuss the influence of
phonon-exciton interaction and electron tunneling effect on the
entangled state. For the real experiment with $t_{e} = 2$ meV, the
concurrence of entangled state is still over $95\%$ at $T = 1$K.

 {\em Acknowledgments.} This work was supported by
National Fundamental Research Program (Grant No. 2009CB929601), also
by National Natural Science Foundation of China (Grant No. 10674128
and 60121503) and the Innovation Funds and \textquotedblleft
Hundreds of Talents\textquotedblright\ program of Chinese Academy of
Sciences and Doctor Foundation of Education Ministry of China (Grant
No. 20060358043).

\end{document}